\documentclass[11pt,twocolumn]{article}

\usepackage[a4paper,margin=18mm,columnsep=7mm]{geometry}
\usepackage{amsmath}
\usepackage{array}
\usepackage{booktabs}
\usepackage{graphicx}
\usepackage{microtype}
\usepackage{tikz}
\usepackage{url}
\usepackage[hidelinks]{hyperref}
\usepackage{xcolor}
\usetikzlibrary{arrows.meta,calc,fit,positioning,shapes.geometric,patterns}

\hypersetup{
  pdftitle={RV565: Cross-Layer Co-Design for Frame-Independent Video Playback on MicroPython-Class ESP32-S3 Systems},
  pdfauthor={Ayushman Bhattacharya, Anwesha Chakraborty},
  pdfsubject={Compressed RGB565 media playback on an ESP32-S3 application platform},
  pdfkeywords={RV565, ESP32-S3, RGB565, MicroPython, DMA, dual-core pipeline, embedded video, Deflate, SPI display}
}

\newcommand{\system}{RV565}
\newcommand{\artifactdoi}{10.5281/zenodo.23543367}

\title{\system: Cross-Layer Co-Design for Frame-Independent\\
Video Playback on MicroPython-Class ESP32-S3 Systems}
\author{Ayushman Bhattacharya\textsuperscript{1} \quad
Anwesha Chakraborty\textsuperscript{2}\\[-0.15em]
{\small \textsuperscript{1}Pollinations AI \quad
\textsuperscript{2}Department of Computer Science and Engineering, JIS University}\\[-0.15em]
{\scriptsize \texttt{ayushman.elixpo@gmail.com} \quad
\texttt{anwesha.elixpo@gmail.com}}}
\date{}

\begin{document}
\maketitle

\begin{abstract}
Video on SPI-attached microcontroller displays is limited less by decoding
than by where each byte lives and which stages run serially.  We present
\system, a decode framework for frame-independent video whose decoded
representation is the panel's native RGB565 format, and evaluate it on an
ESP32-S3 with two ten-second windows of an openly licensed film.  A serial
path---inflate, then scale into internal-memory stripes overlapped with SPI
direct-memory-access (DMA) transfers---meets the 41.7~ms deadline on the
lighter window but misses all 7,200 frames of the heavier one.  \system's
dual-core pipeline inflates the next frame on the second core during the
transfer and sustains 24~FPS with zero misses and 10.3~ms of slack on both
windows; frame time then equals the display transfer, which runs at 98.1\% of
the 40~MHz wire-time bound.  Because frames are independent, inflation time
is linear in the compressed length stored in the container (pooled
$r^2=0.9987$): a model fitted on one window predicts the other within
0.04~ms and identifies from metadata alone which window the serial path
cannot schedule.  We show why Deflate cannot exploit co-located temporal
redundancy in RGB565 frames larger than 128$\times$128 pixels and measure the
price: a shared Deflate stream changes size by at most 1.1\%, whereas an
explicit XOR delta would save 46--80\%.  Inside OreoOS, a MicroPython
application operating system, firmware inflation alone reaches 12.2~FPS;
stage timing attributes the 80.4~ms frame time to serialized filesystem
input, external-RAM placement, and blocking presentation, with 0.8~ms of
interpreter overhead.  Firmware, measurements, and analysis scripts are
released.
\end{abstract}

\section{Introduction}

Video playback on a microcontroller is not solely a decoding problem.  A
complete frame must be fetched, decoded, transformed to the panel's pixel
layout, and transferred before its presentation deadline.  At the same time,
an interactive device must continue polling buttons and servicing operating
system work.  These constraints are especially visible on SPI-connected
displays, for which transfer time consumes a substantial and predictable part
of every frame period.

General-purpose video formats optimise compression ratio, interoperability,
and visual quality.  They commonly introduce container parsing, colour-space
conversion, variable decoder state, and inter-frame dependencies.  Those
features are valuable on larger systems, but can make latency and memory use
difficult to bound on a microcontroller.  Raw RGB565 has the opposite
trade-off: display transfer is simple and deterministic, while storage and
input bandwidth scale directly with the number of pixels.

Existing approaches can be organised along three architectural axes.  The
first is \emph{where decode complexity resides}: display-native raw
playback~\cite{microchip2012video} moves all work to the host and pays in
storage, whereas transform codecs such as JPEG~\cite{wallace1991jpeg} and its
embedded decoders~\cite{chan2021tjpgd,espressif2026newjpeg} place inverse
transforms and colour conversion on the device.  The second is
\emph{inter-frame dependence}: motion-compensated codecs such as
H.264/AVC~\cite{wiegand2003overview} reach high compression through reference
pictures, while intra-only streams such as MJPEG decode every frame in
isolation; software decoders for dependent streams on general-purpose
processors~\cite{lee1995realtime} and media processors~\cite{mattavelli1998implementing}
had to predict and schedule variable decode complexity to stay real-time.
The third is \emph{the distance between the decoded format and the panel
format}: decoders that emit RGB888 or YUV require a conversion pass before an
RGB565 panel can be driven, while byte-oriented lossless formats such as
QOI~\cite{szablewski2022qoi} and LZ4~\cite{collet2024lz4} trade compression
ratio for decode speed but are not tied to a display format.  Integrated
frameworks~\cite{espressif2026player} span several of these points.  On a
microcontroller, each axis maps onto a hardware resource: device decode work
consumes CPU cycles during the frame period, inter-frame state consumes
scarce internal SRAM, and format conversion adds a full pass over the frame
through the memory hierarchy.

\system{} occupies a deliberately narrow point in this space: host-side
resampling and colour quantisation, intra-only frames, and lossless Deflate
payloads~\cite{deutsch1996deflate} whose decoded bytes are already the panel's
RGB565 format.  Its decode framework comprises the container and its
validation contract, a dual-core pipeline that places inflation, scaling, and
DMA transfer on separate execution resources, an absolute-deadline scheduler,
and a cost model that predicts schedulability from container metadata.  The
entropy-decoding primitive is the standard \texttt{tinfl} inflater; we do not
propose a new entropy coder.  This work asks where the frame budget goes for
such a stream on an ESP32-S3, both on bare firmware and inside
OreoOS~\cite{bhattacharya2026oreo}, a MicroPython~\cite{micropython}
application operating system.

Our contributions to the study of embedded media architectures are as
follows.  We define a frame-independent, display-native container with a
host- and device-side validation contract
(Sections~\ref{sec:container}--\ref{sec:validation}).  We design a dual-core
decode pipeline and evaluate it against the serial path as a controlled
ablation on two workloads: it removes inflation from the critical path,
holding 24~FPS with zero misses on content that the serial path cannot
schedule, and reduces frame time to the display transfer, which runs within
1.9\% of the wire-time bound (Section~\ref{sec:bus}).  We show that
intra-only Deflate makes decode cost a linear function of the payload length
already stored in the container, so that the complexity prediction earlier
real-time decoders had to estimate~\cite{mattavelli1998implementing} becomes
available from metadata before playback; the model transfers between windows
and yields an admission threshold that classifies every frame correctly
(Section~\ref{sec:model}).  We derive the condition under which Deflate's
match window cannot exploit co-located temporal redundancy and quantify the
real price of frame independence (Section~\ref{sec:independence}).  Finally,
we provide a stage-resolved characterisation of interpreter-hosted playback
showing that the gap between bare firmware and an application OS is caused by
serialization and memory placement rather than interpreter overhead
(Section~\ref{sec:gap}).  All firmware, measurements, and analysis scripts
are public~\cite{bhattacharya2026oreo}.

\section{RGB565 as the Design Starting Point}

RGB565 encodes one pixel in 16 bits: five bits represent red, six represent
green, and five represent blue.  Green receives the additional bit because
human vision is comparatively sensitive to luminance detail in that portion
of the spectrum.  A pixel can be written as

\begin{equation}
p=(r_5 \ll 11)\;|\;(g_6 \ll 5)\;|\;b_5,
\label{eq:pixel}
\end{equation}

where $r_5$, $g_6$, and $b_5$ are quantised colour components.  \system{}
stores the 16-bit value of Eq.~\eqref{eq:pixel} in the big-endian byte order
that the panel controller expects on its serial interface, so a decoded frame
can move toward the display without an intermediate RGB888 or YUV conversion.

This choice simplifies the device but exposes the system constraint.  A frame
contains $W\times H$ pixels and therefore occupies

\begin{equation}
B_{frame}=2WH
\label{eq:frame}
\end{equation}

bytes before compression (Eq.~\eqref{eq:frame}).  Frame rate multiplies that
value into a continuous storage and memory-bandwidth demand, while the final
320$\times$240 image must still cross the SPI bus.  \system{} consequently
compresses what is stored and read, not what the display ultimately receives.
Every successfully decoded frame has a deterministic size and colour layout,
which keeps validation, scaling, and presentation straightforward.

RGB565 also defines the quality boundary.  Deflate is lossless with respect to
the encoded bytes, but conversion from the original colour space and spatial
resizing are not.  Any loss in \system{} therefore arises from resampling and
RGB565 quantisation, never from compression.  \system{} is a media
architecture built around a native pixel representation, not a claim that
16-bit colour is visually lossless relative to the source.

\section{From a Gallery Feature to a Systems Problem}

The project began with a modest goal: Gallery already displayed still images,
so video appeared to be a sequence of images plus a clock.  The first player
followed that model literally.  It read a frame, decoded it in MicroPython,
scaled pixels in Python loops, copied the result into the framebuffer, and
asked the normal display driver to present it.  The result was technically a
video player, but not an interactive one.  A frame occupied the interpreter
for so long that playback advanced visibly rather than continuously, and
button events waited behind media work.

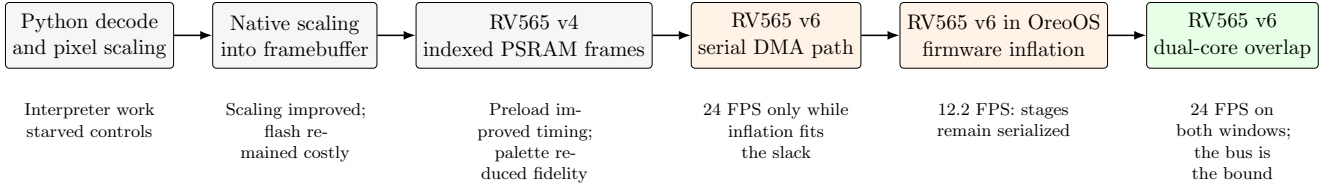
\begin{figure*}[t]
\centering
\resizebox{\textwidth}{!}{%
\begin{tikzpicture}[
  stage/.style={draw, rounded corners=2pt, minimum width=30mm,
                minimum height=12mm, align=center, fill=gray!7},
  lesson/.style={font=\footnotesize, align=center, text width=29mm},
  arrow/.style={-{Latex[length=2.2mm]}, thick},
  node distance=7mm
]
  \node[stage] (python) {Python decode\\and pixel scaling};
  \node[stage, right=of python] (native) {Native scaling\\into framebuffer};
  \node[stage, right=of native] (v4) {\system{} v4\\indexed PSRAM frames};
  \node[stage, right=of v4, fill=orange!10] (serial) {\system{} v6\\serial DMA path};
  \node[stage, right=of serial, fill=orange!10] (os) {\system{} v6 in OreoOS\\firmware inflation};
  \node[stage, right=of os, fill=green!10] (overlap) {\system{} v6\\dual-core overlap};

  \draw[arrow] (python) -- (native);
  \draw[arrow] (native) -- (v4);
  \draw[arrow] (v4) -- (serial);
  \draw[arrow] (serial) -- (os);
  \draw[arrow] (os) -- (overlap);

  \node[lesson, below=5mm of python] {Interpreter work\\starved controls};
  \node[lesson, below=5mm of native] {Scaling improved;\\flash remained costly};
  \node[lesson, below=5mm of v4] {Preload improved timing;\\palette reduced fidelity};
  \node[lesson, below=5mm of serial] {24 FPS only while\\inflation fits the slack};
  \node[lesson, below=5mm of os] {12.2 FPS: stages\\remain serialized};
  \node[lesson, below=5mm of overlap] {24 FPS on both windows;\\the bus is the bound};
\end{tikzpicture}}
\caption{\system{} evolved by following the bottleneck exposed by each
prototype.  The first three stages are qualitative history; the three
\system{} v6 stages are measured in Section~\ref{sec:evaluation}.  Bringing
the dual-core path into OreoOS remains future work.}
\label{fig:evolution}
\end{figure*}

That failure changed the question.  The problem was no longer how to parse a
video file; it was how to move one complete frame through storage, memory,
conversion, and the display bus before the next deadline while still returning
to the operating-system loop.  Each later version removed one bottleneck from
that path (Figure~\ref{fig:evolution}).  The first three prototypes were not
instrumented, so their history is qualitative; the three \system{} v6 paths
are evaluated quantitatively in Section~\ref{sec:evaluation}.

\subsection{Moving pixels out of Python}

The first correction was a native scaler that wrote RGB565 pixels directly
into the display framebuffer.  This removed nested Python loops and restored
much of the frame budget, but it also revealed the next cost: repeatedly
reading frame-sized payloads from the filesystem.  Optimising the computation
had merely exposed the storage path.

\system{} v4 responded by moving expensive colour analysis to the host.  Each
frame carried an adaptive RGB565 palette and an 8-bit index plane.  Gallery
loaded those independent frame blocks into PSRAM incrementally and the native
module expanded them directly into the output framebuffer.  Blocks were used
instead of one monolithic allocation because a long-running UI can have ample
total PSRAM yet lack one sufficiently large contiguous region.  The design
removed the per-frame filesystem read, but a per-frame palette placed an
avoidable ceiling on image fidelity.

\subsection{Separating the hardware limit from the OS cost}

\system{} v6 returned to full-colour RGB565 frames and compressed each one
independently.  To learn whether the hardware could sustain the intended
playback rate, we temporarily removed the operating system and built a
standalone ESP-IDF experiment that inflates each frame and then transfers the
scaled output through two internal-memory stripes with SPI2 GDMA.  This serial
path meets the deadline only while inflation fits in the slack left by the
display transfer, which depends on content (Section~\ref{sec:standalone-results}).

Reintroducing the operating system made a second boundary visible.  We moved
Deflate decoding below the Python boundary into a firmware-resident module,
but kept Gallery's existing filesystem input, framebuffer scaling, and display
driver; that integrated path reaches 12.170~FPS, and Section~\ref{sec:gap}
decomposes the shortfall stage by stage.  Measurements of the serial path then
showed the CPU idle for most of each transfer, which led to the dual-core
pipeline of Section~\ref{sec:overlap}.  This history matters to the method:
the architecture was derived by measuring and relocating the critical path one
layer at a time.

\section{Background and Previous Work}

\system{} combines established ideas---display-native pixels, independent
compression, external memory, and DMA-backed transfers---under the constraints
of an interactive MicroPython system.  Table~\ref{tab:literature} compares the
approaches introduced above along the axes of the taxonomy; this section
discusses each family and separates its foundations from the contribution of
\system{}.

\begin{table*}[t]
\centering
\caption{Comparison of video and image representations relevant to
SPI-display microcontrollers, organised by the taxonomy of the introduction.
``Inter-frame state'' is decoder state carried between frames; ``device
work'' is the dominant per-frame operation on the decoding processor.}
\label{tab:literature}
\small
\setlength{\tabcolsep}{4pt}
\begin{tabular}{@{}>{\raggedright\arraybackslash}p{0.18\textwidth}%
                >{\raggedright\arraybackslash}p{0.18\textwidth}%
                >{\raggedright\arraybackslash}p{0.10\textwidth}%
                >{\raggedright\arraybackslash}p{0.18\textwidth}%
                >{\raggedright\arraybackslash}p{0.11\textwidth}%
                >{\raggedright\arraybackslash}p{0.14\textwidth}@{}}
\toprule
Approach & Representation & Inter-frame state & Device work per frame & Random access & Target platform \\
\midrule
Raw RGB565 playback~\cite{microchip2012video} & Uncompressed display-native frames & None & Copy to panel & Fixed offsets & PIC MCU \\
JPEG / TJpgDec~\cite{wallace1991jpeg,chan2021tjpgd} & Baseline JPEG; RGB565 output option & None & Huffman, IDCT, colour conversion & Per image & Low-memory MCU \\
ESP\_NEW\_JPEG~\cite{espressif2026newjpeg} & JPEG with SIMD acceleration & None & Huffman, IDCT, colour conversion & Per image & ESP32-S3 \\
ESP Player~\cite{espressif2026player} & MJPEG or H.264 in a container & H.264 only & Demux, decode, render, A/V sync & Container seek & Espressif SoCs \\
H.264/AVC~\cite{wiegand2003overview} & Motion-compensated transform coding & Reference pictures & Prediction, transform, filtering & From key frames & Hardware or large CPU \\
Software MPEG~\cite{lee1995realtime} & MPEG-1 & Reference pictures & Software decode with media instructions & From key frames & Workstation RISC CPU \\
Complexity prediction~\cite{mattavelli1998implementing} & MPEG with predicted decode cost & Reference pictures & Decode under a predicted schedule & From key frames & Media processor \\
QOI~\cite{szablewski2022qoi} & Byte-oriented lossless image & None & Run, index and delta ops & Per image & General \\
LZ4 frame~\cite{collet2024lz4} & LZ77 without entropy stage & Within stream & Byte-copy decoding & None in frame format & General \\
\midrule
\system{} v6 (this work) & Display-native RGB565, per-frame Deflate & None & Inflate on a second core; scaling overlapped with DMA & $O(1)$ via length index & ESP32-S3, bare or MicroPython \\
\bottomrule
\end{tabular}
\end{table*}

\subsection{Display-native playback}

Microcontroller video demonstrations have long stored frames in a form close
to the display representation.  Microchip's AN1415, for example, describes
RGB565 video playback from removable storage~\cite{microchip2012video}.  Such
designs minimise device-side transformation but trade storage capacity and
input bandwidth for simple execution.  \system{} retains a display-native
decoded frame while compressing each frame independently to reduce storage
and input traffic.

\subsection{Transform codecs and embedded decoders}

JPEG is a widely deployed still-image coding standard~\cite{wallace1991jpeg},
and concatenated JPEG images form a practical MJPEG stream.  TJpgDec is a
baseline JPEG decompressor designed for very small memory footprints, with an
RGB565 output option~\cite{chan2021tjpgd}; Espressif's ESP\_NEW\_JPEG library
uses the ESP32-S3's SIMD instructions to accelerate JPEG encoding and
decoding~\cite{espressif2026newjpeg}.  Espressif's multimedia framework
supports MJPEG and H.264 playback, demultiplexing, rendering, seeking, and
audio/video synchronisation~\cite{espressif2026player}, and is therefore an
important implementation baseline.  These decoders spend device cycles on
entropy decoding, inverse transforms, and colour conversion; \system{} moves
the transform and quantisation to the host and leaves only inflation.

\subsection{Inter-frame codecs and decode-cost prediction}

H.264/AVC achieves high compression with motion-compensated prediction from
reference pictures~\cite{wiegand2003overview}.  Software decoding of such
streams was first made real-time on general-purpose processors with media
instructions~\cite{lee1995realtime}.  Because the decode cost of dependent
frames varies with content and reference structure, Mattavelli et al.\
predicted per-frame complexity to schedule decoding on media
processors~\cite{mattavelli1998implementing}.  \system{} reaches the same
scheduling goal from the opposite direction: with intra-only, single-codec
frames, decode cost becomes a measured linear function of a length that the
container already stores (Section~\ref{sec:model}).

\subsection{Lossless compression and independent frames}

Deflate combines the dictionary construction introduced by Ziv and
Lempel~\cite{ziv1977universal} with Huffman's minimum-redundancy prefix
codes~\cite{huffman1952method}; RFC~1951 defines their use in the Deflate
bitstream, including a maximum back-reference distance of
32,768~bytes~\cite{deutsch1996deflate}.  The zlib wrapper specifies framing
and integrity metadata~\cite{deutsch1996zlib}.  Both \system{} decoders use
the \texttt{tinfl} inflater from miniz~\cite{miniz}, which the ESP32-S3 also
provides in ROM.  Faster byte-oriented alternatives exist: LZ4 omits the
entropy stage entirely~\cite{collet2024lz4}, and QOI encodes images with a
small set of run, index, and delta operations~\cite{szablewski2022qoi}.
\system{} uses Deflate because a ROM decoder is already present on the target
and because its entropy stage yields smaller payloads, which matters on the
filesystem path (Section~\ref{sec:gap}).  Compressing each frame separately
resets the dictionary at every boundary; this bounds corruption, enables
direct seeking, and avoids reconstructing a dependency chain after a skipped
frame, and its compression cost is measured in
Section~\ref{sec:independence}.

\subsection{ESP32-S3 memory and display I/O}

The ESP32-S3 combines dual Xtensa LX7 cores with internal SRAM, optional
octal PSRAM, memory-mapped flash behind a cache, and a general-purpose DMA
controller shared by peripherals~\cite{espressif2023datasheet,espressif2024trm}.
ESP-IDF exposes queued SPI transactions backed by that
DMA~\cite{espressif2026spi}.  These facilities do not themselves create a
video architecture: the relevant design question is how decoded data is
partitioned across this memory hierarchy and how work is placed on the two
cores and the DMA engine so that the panel transfer is never starved.

\section{The \system{} Method}

The method follows the path of a frame from offline preparation to visible LCD
output.  Its organising principle is to move expensive, unconstrained work to
the host while keeping every device-side stage bounded, observable, and, where
possible, off the critical path.

\subsection{Design goals}

The architecture targets four properties: predictable per-frame work,
display-native colour, bounded device memory, and responsive controls.  It
assumes offline access to substantially more compute than is available on the
badge.  Expensive resize and format decisions are therefore moved to the host;
the device performs only parsing, inflation, nearest-neighbour scaling, and
display transfer.

\subsection{Why frame independence}

\system{} v6 makes every compressed payload a complete frame rather than a
delta from an earlier image.  A decoder can therefore start from any indexed
frame with only the global header and that frame's four-byte length prefix.
This property is useful beyond seeking: looping does not require decoder
history, a missed deadline can be handled by advancing the frame index, and a
damaged payload cannot alter the dictionary used by later frames.  After one
linear scan builds the offset table in $O(N)$ time, lookup is $O(1)$.

Within Deflate, independence costs less than intuition suggests.  When
consecutive frames are concatenated into one stream, the co-located pixel of
the previous frame lies exactly $2WH$ bytes behind the current position by
Eq.~\eqref{eq:frame}.  Because a Deflate back-reference may reach at most
32,768~bytes~\cite{deutsch1996deflate}, the stream cannot reference that pixel
whenever

\begin{equation}
2WH > 32{,}768 \quad\Longleftrightarrow\quad WH > 16{,}384,
\label{eq:window}
\end{equation}

that is, for any RGB565 frame larger than 128$\times$128 pixels.  Above that
size, a shared stream can still reuse statistics and spatially displaced
content near the end of the previous frame, but not static co-located content,
which is a principal source of temporal redundancy in video.  Exploiting that
redundancy requires explicit prediction from a reference frame, as in
inter-frame codecs~\cite{wiegand2003overview}, at the cost of reference state
and catch-up decoding after a seek.  For the 180$\times$135 frames evaluated
here, $2WH=48{,}600$~bytes; Section~\ref{sec:independence} measures both
sides of this trade-off.

\subsection{End-to-end architecture}

\begin{figure*}[t]
\centering
\resizebox{0.98\textwidth}{!}{%
\begin{tikzpicture}[
  font=\small,
  node distance=7mm and 9mm,
  block/.style={draw, rounded corners=2pt, minimum height=14mm,
                minimum width=25mm, align=center, fill=gray!7},
  store/.style={draw, cylinder, shape border rotate=90, aspect=0.25,
                minimum height=17mm, minimum width=19mm, align=center,
                fill=blue!6},
  fast/.style={draw, rounded corners=2pt, minimum height=14mm,
               minimum width=24mm, align=center, fill=orange!13},
  core/.style={draw, rounded corners=2pt, minimum height=14mm,
               minimum width=25mm, align=center, fill=green!10},
  control/.style={draw, dashed, rounded corners=2pt, minimum height=13mm,
                  minimum width=25mm, align=center, fill=green!7},
  flow/.style={-{Latex[length=2.2mm]}, thick},
  ctl/.style={-{Latex[length=2mm]}, dashed}
]
  \node[block] (source) {Source video};
  \node[block, right=of source] (host) {Host crop, resize,\\frame-rate conversion};
  \node[block, right=of host] (rgb) {RGB888 $\rightarrow$\\RGB565BE};
  \node[block, right=of rgb] (deflate) {Independent\\zlib level 1 frame};
  \node[store, right=of deflate] (container) {\system{} v6\\container};

  \node[store, below=44mm of container] (storage) {Memory-mapped\\flash};
  \node[block, left=of storage] (parser) {Header validation\\and frame index};
  \node[core, left=of parser] (inflate) {\texttt{tinfl} inflate\\core 1 (overlap)};
  \node[store, left=of inflate] (frame) {Decoded frames\\$k$, $k{+}1$ in SRAM};
  \node[block, left=of frame] (scale) {Nearest-neighbour\\stripe scaler, core 0};
  \node[fast, anchor=east] (stripea) at ($(scale.west)+(-12mm,11mm)$)
       {Stripe $n$\\internal SRAM A};
  \node[fast, anchor=east] (stripeb) at ($(scale.west)+(-12mm,-11mm)$)
       {Stripe $n{+}1$\\internal SRAM B};
  \node[fast, anchor=east] (dma) at ($(stripea.west |- scale.west)+(-12mm,0)$)
       {SPI2 GDMA\\queued transfer};
  \node[block, left=of dma] (lcd) {320$\times$240\\RGB565 LCD};

  \node[control] (scheduler) at ($(parser)+(0,-36mm)$)
       {Absolute-deadline loop\\prefetch $k{+}1$; skip when late};

  \draw[flow] (source) -- (host);
  \draw[flow] (host) -- (rgb);
  \draw[flow] (rgb) -- (deflate);
  \draw[flow] (deflate) -- (container);
  \draw[flow] (container) -- (storage);
  \draw[flow] (storage) -- (parser);
  \draw[flow] (parser) -- (inflate);
  \draw[flow] (inflate) -- (frame);
  \draw[flow] (frame) -- (scale);
  \draw[flow] (scale.west) -- ++(-6mm,0) |- (stripea.east);
  \draw[flow] (scale.west) -- ++(-6mm,0) |- (stripeb.east);
  \draw[flow] (stripea.west) -- ++(-6mm,0) |- (dma.east);
  \draw[flow] (stripeb.west) -- ++(-6mm,0) |- (dma.east);
  \draw[flow] (dma) -- (lcd);
  \draw[ctl] (scheduler) -- (parser);
  \draw[ctl] (scheduler.west) -| (inflate.south);
  \draw[ctl] (scheduler.south) -- ++(0,-4mm) -| (dma.south);

  \node[draw, dashed, fit=(source)(host)(rgb)(deflate)(container),
        label=above:{\bfseries Host-side preparation}, inner sep=4mm] {};
  \node[draw, dashed, fit=(storage)(parser)(inflate)(frame)(scale)(stripea)(stripeb)(dma)(lcd),
        label={[anchor=south west]north west:{\bfseries ESP32-S3 standalone data path}},
        inner sep=4mm] {};
\end{tikzpicture}}
\caption{The \system{} standalone architecture.  Frame independence permits
indexed seeking and local recovery.  Alternating DMA-capable stripes let core
0 scale while GDMA transfers; in the overlapped configuration core 1 inflates
the next frame into a second buffer at the same time.  Dashed arrows carry
scheduling.  The OreoOS integration replaces the lower path with the serial
stages of Figure~\ref{fig:timeline}.}
\label{fig:architecture}
\end{figure*}
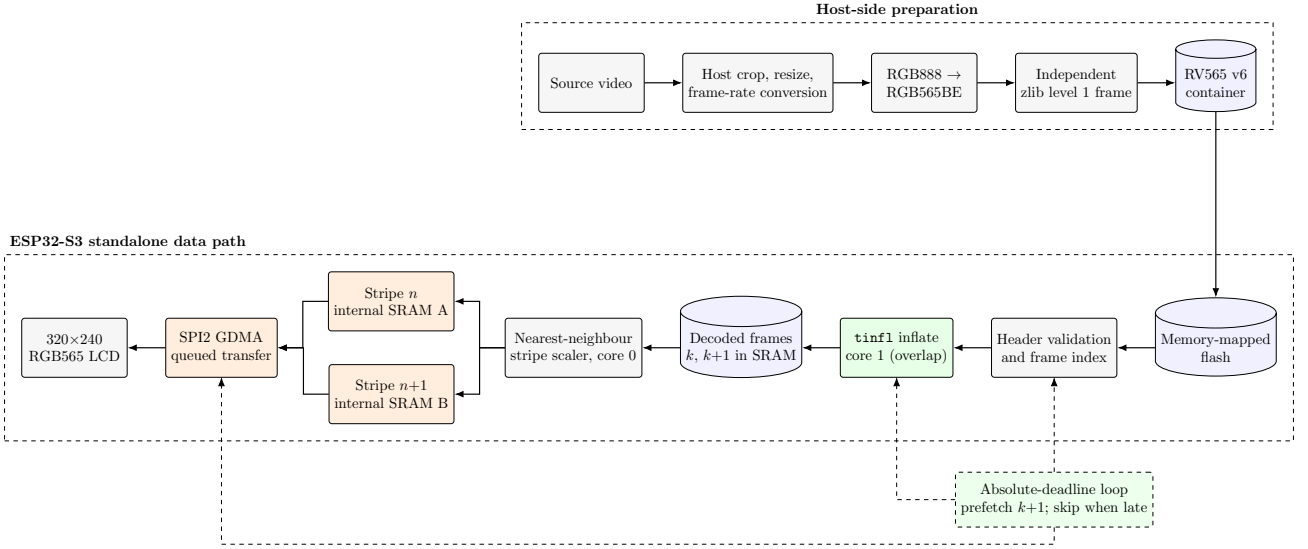

Figure~\ref{fig:architecture} shows the standalone data path.  The container
is linked into the application image and read through the memory-mapped flash
cache, so the standalone path has no separate filesystem read.  Decoded frames
(48,600~bytes each) and two 24-row output stripes (15,360~bytes each) reside
in internal SRAM; the standalone firmware does not enable the external-RAM
allocator.  While GDMA transfers stripe $n$, the CPU scales stripe $n+1$ into
the alternate buffer.  In the serial configuration a single core inflates a
frame and then transfers it; in the overlapped configuration the second core
inflates the next frame into a second buffer during the transfer.

\begin{figure*}[!tp]
\centering
\includegraphics[width=0.94\textwidth]{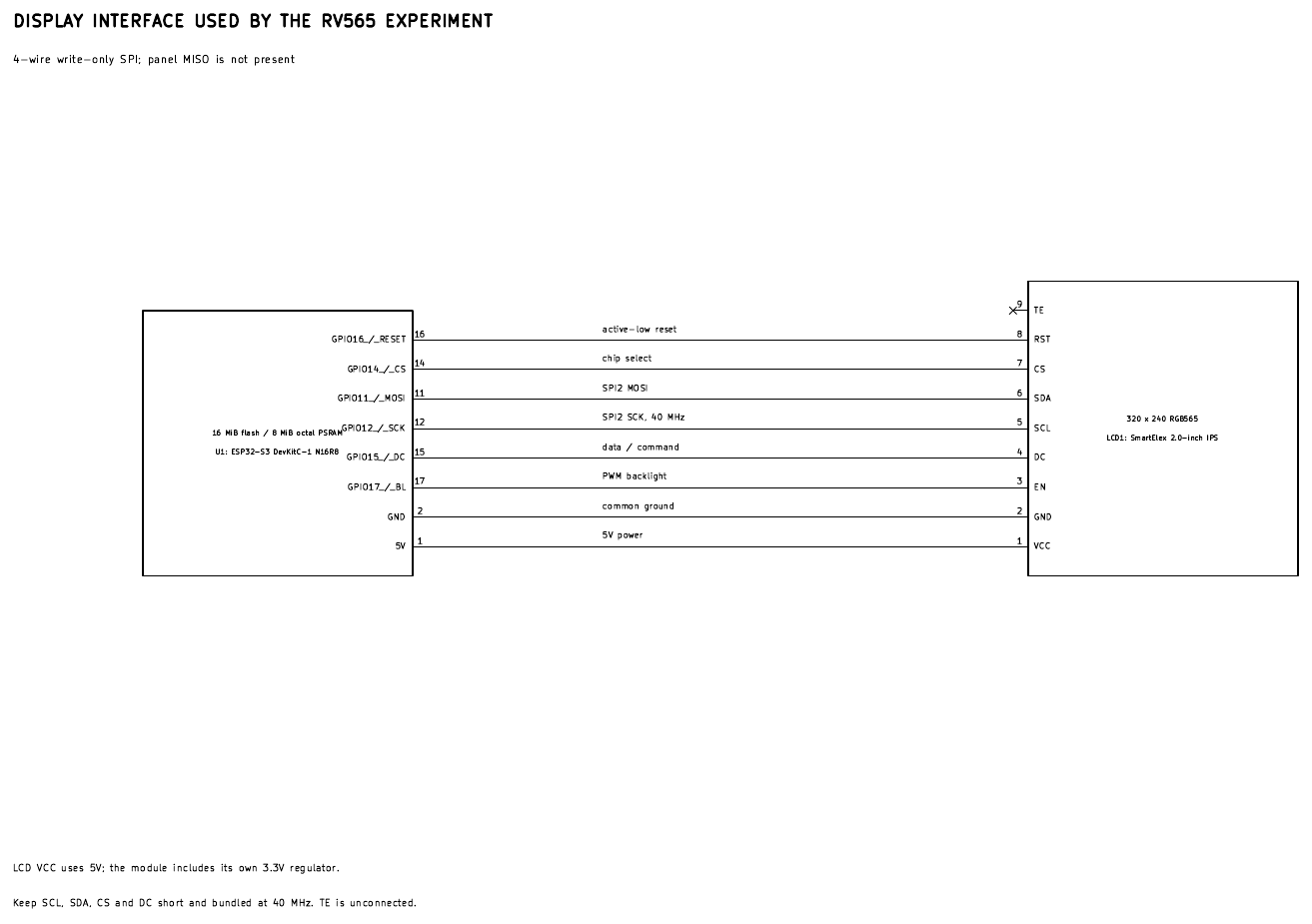}
\\[-1mm]\textbf{(a)} LCD connection and signal path
\\[2mm]
\includegraphics[width=0.94\textwidth]{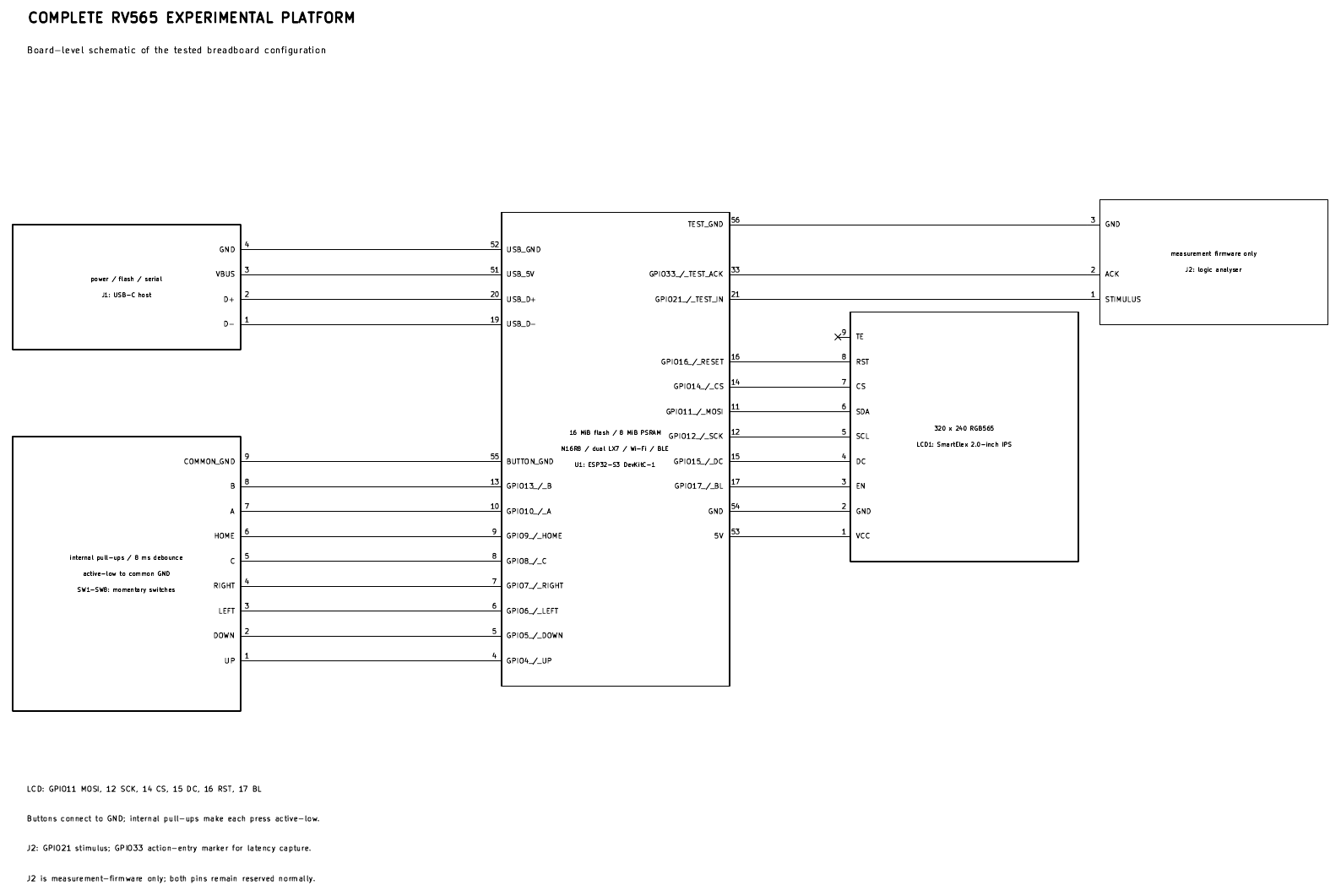}
\\[-1mm]\textbf{(b)} Complete experimental circuit
\caption{Board-level schematics of the \system{} test platform.  Panel (a)
shows the 40~MHz write-only ST7789 interface.  Panel (b) shows the complete
video and interaction test path; GPIO21 and GPIO33 are reserved for stimulus
and acknowledgement markers in future input-latency firmware.}
\label{fig:hardware-platform}
\end{figure*}

Figure~\ref{fig:hardware-platform} documents the hardware boundary used by the
measurements; the schematics were drawn in KiCad~\cite{kicad}.  The first
panel isolates the
LCD clock, data, command/data, reset, chip-select, and backlight paths.  The
second places that display beside the DevKit's flash/PSRAM configuration, USB
connection, active-low controls, and the header reserved for future
input-latency tests.

\subsection{Container format}
\label{sec:container}

A \system{} v6 file begins with a 12-byte header containing a magic
identifier, format version, source width and height, frame rate, flags, and
frame count.  Each subsequent frame is stored as a little-endian 32-bit
compressed length followed by one zlib-wrapped Deflate payload.  Successful
inflation produces exactly $w\times h\times2$ bytes in RGB565 big-endian
order (Table~\ref{tab:format}).  The flags byte is currently zero.

\begin{table}[t]
\centering
\caption{\system{} v6 file layout. Multi-byte integers are little-endian;
decoded pixels are RGB565 big-endian.}
\label{tab:format}
\resizebox{\columnwidth}{!}{%
\begin{tabular}{rll}
\toprule
Offset & Size & Field \\
\midrule
0 & 3 & Magic, \texttt{RV5} \\
3 & 1 & Version, \texttt{6} \\
4 & 2 & Source width \\
6 & 2 & Source height \\
8 & 1 & Frame rate \\
9 & 1 & Flags (zero) \\
10 & 2 & Frame count \\
12 & 4 & Compressed length, frame 0 \\
16 & variable & zlib/Deflate payload, frame 0 \\
$\cdots$ & $\cdots$ & Repeated length and payload \\
\bottomrule
\end{tabular}
}
\end{table}

The length prefix permits a frame-offset table to be constructed in one scan.
Seeking then changes the compressed input position directly and does not
decode preceding frames.  Gallery builds the table lazily on the first seek;
the standalone firmware builds pointer and length tables (1,920~bytes for 240
frames) when the clip is opened.  Because lengths are known before any
payload is decoded, they also predict decode cost
(Section~\ref{sec:model}).

\subsection{Validation and failure recovery}
\label{sec:validation}

Media is checked at two boundaries.  Before transfer, the host ingest tool
decodes every compressed frame and verifies the magic, version, dimensions,
frame rate, frame count, exact RGB565 output length, complete payload
coverage, and absence of trailing data.  It then compares the encoded
artifact---not the source file---with both a media budget and the free space
reported by the badge.  An interrupted or malformed conversion is consequently
rejected before it can become Gallery input.

Device validation is repeated because stored media cannot be assumed to remain
trusted.  Dimensions must fit the 320$\times$240 panel, frame rate is limited
to 1--30~FPS, and frame count and compressed lengths must be nonzero and fit
the file and the allocated input buffer.  The standalone firmware
additionally rejects truncated tables and trailing bytes.  Inflation succeeds
only when it produces exactly $2wh$ bytes, so a partial decode is never
presented as a new frame.  Gallery currently fails closed when retrying the
frame after a rewind also fails.  Independence would permit a future policy
to skip a corrupt payload and resume at the next valid boundary; that policy
is not implemented.

\subsection{Host encoding}

The encoder uses FFmpeg~\cite{ffmpeg} to seek to the
start of the requested window, resample the source to the target frame rate,
resize it with a Lanczos filter~\cite{duchon1979lanczos} while preserving
aspect ratio, crop to the target dimensions, and convert to RGB565BE
(Eq.~\eqref{eq:pixel}).  Every frame is then compressed separately into a
zlib stream~\cite{deutsch1996zlib} at level~1, the fastest setting.  Encoder
complexity is thus irrelevant to device timing, and the level is a deliberate
choice of fast compression rather than minimum size
(Section~\ref{sec:independence}).

\subsection{Serial decode and stripe pipeline}

The standalone decoder reuses all large allocations, which are made
with ESP-IDF's capability-based allocator so that buffers can be restricted
to DMA-capable internal memory~\cite{espressif2026memalloc}.  In the serial
configuration it inflates one independent payload directly from
memory-mapped flash into the decoded frame buffer, then scales output rows
into one of two internal-memory stripes of 24 rows each, ten stripes per
frame.  The chip select is held asserted across the panel's RAMWR command and
all stripes.  The first stripe is scaled and queued as an SPI transaction
serviced by GDMA~\cite{espressif2026spi,espressif2024trm}; while it is
transferred, the CPU fills the other stripe; a wait for the completed
transaction precedes reuse of a buffer, after which the roles alternate.
Serial frame work is therefore

\begin{equation}
T_{work}^{ser}=T_{inflate}+T_{lcd},\qquad
T_{lcd}\approx T_{wire}+T_{scale}^{(1)}+\epsilon,
\label{eq:work}
\end{equation}

where $T_{wire}$ is the SPI wire time, $T_{scale}^{(1)}$ the unoverlapped
first stripe, and $\epsilon$ command and transaction overhead.  The remaining
nine stripes are scaled under DMA.  For a 320$\times$240 RGB565 frame on a
40~MHz single-lane bus,

\begin{equation}
T_{wire}=\frac{320\cdot240\cdot16\ \text{bit}}{40\ \text{MHz}}=30.72\ \text{ms},
\label{eq:wire}
\end{equation}

which bounds full-screen presentation at 32.55~FPS regardless of decoding.
Playback meets its target only when the upper tail of $T_{work}$, not merely
the mean, is below the period; the evaluation reports percentiles and the
worst observed value.

\subsection{Dual-core overlapped decode}
\label{sec:overlap}

The serial path leaves the CPU waiting for DMA for most of each transfer.
The overlapped configuration exploits this by moving inflation to the second
core.  A worker task pinned to core 1 owns inflation and communicates with
the playback loop on core 0 through two single-entry queues.  Before
transferring frame $k$, core 0 requests frame $k+1$ into the alternate
decoded buffer; it then scales and transfers frame $k$ while core 1 inflates,
and finally waits for the worker's completion message before swapping the
buffers.  Frame work becomes

\begin{equation}
T_{work}^{ovl}=\max\bigl(T_{lcd},\,T_{inflate}^{(k+1)}\bigr)+\epsilon_{sync},
\label{eq:overlap}
\end{equation}

so inflation leaves the critical path whenever it is shorter than the
transfer.  The prefetch is speculative: if the scheduler skips frames, the
prefetched frame is stale and the required frame is inflated on the critical
path before playback resumes.  The configuration costs a second 48,600-byte
decoded buffer and a 24~KiB worker stack, because \texttt{tinfl} keeps its
decompressor state on the stack.  Both configurations are built from the same
source and selected at compile time, and the firmware reports the active
configuration with every run, so the comparison in Section~\ref{sec:bus} is a
controlled ablation.

\subsection{OreoOS integration}

\begin{figure*}[t]
\centering
\begin{tikzpicture}[x=1.62mm, y=1mm, font=\scriptsize]
  \draw[red!70!black, densely dashed, thick] (41.667,-1) -- (41.667,37)
    node[above] {41.67 ms deadline (24 FPS)};
  \node[anchor=east, align=right, font=\footnotesize] at (-1,28) {Standalone\\serial};
  \draw[fill=blue!25] (0,24) rectangle (8.910,32);
  \node[align=center] at (4.45,28) {inflate\\8.9};
  \draw[fill=orange!25] (8.910,24) rectangle (40.222,32);
  \fill[pattern=north east lines, pattern color=orange!70!black]
    (8.910,24) rectangle (11.508,32);
  \node[align=center] at (25.5,28) {LCD 31.3\\wire 30.72; scaling under DMA};
  \node[anchor=east] at (40.0,34) {\textbf{40.2}};
  \node[anchor=east, align=right, font=\footnotesize] at (-1,16) {Standalone\\dual-core};
  \draw[fill=orange!25] (0,16) rectangle (31.316,21);
  \fill[pattern=north east lines, pattern color=orange!70!black]
    (0,16) rectangle (2.598,21);
  \node at (16.9,18.5) {core 0: LCD 31.3};
  \draw[fill=blue!25] (0,11) rectangle (8.902,16);
  \node at (4.45,13.5) {8.9};
  \node[anchor=west] at (9.3,13.5) {core 1: inflate frame $k{+}1$};
  \node[anchor=west] at (31.8,18.5) {\textbf{31.3}};
  \node[anchor=east, align=right, font=\footnotesize] at (-1,4) {OreoOS\\Gallery};
  \draw[fill=gray!25] (0,0) rectangle (27.631,8);
  \node[align=center] at (13.8,4) {filesystem read\\27.6};
  \draw[fill=blue!25] (27.631,0) rectangle (37.156,8);
  \node[align=center] at (32.39,4) {inflate\\9.5};
  \draw[fill=green!20] (37.156,0) rectangle (46.256,8);
  \node[align=center] at (41.7,4) {scale +\\overlay 9.1};
  \draw[fill=orange!25] (46.256,0) rectangle (79.573,8);
  \node[align=center] at (62.9,4) {present 33.3\\(blocking SPI)};
  \draw[fill=black!50] (79.573,0) rectangle (80.400,8);
  \node[anchor=west] at (80.9,4) {\textbf{80.4}};
  \draw (0,-2) -- (84,-2);
  \foreach \t in {0,10,20,30,40,50,60,70,80} {
    \draw (\t,-2) -- (\t,-3); \node[below] at (\t,-3) {\t};
  }
  \node[below] at (42,-6.5) {mean frame work on W90 (ms)};
\end{tikzpicture}
\caption{Mean per-frame critical path on W90, drawn to scale from the
measured stage times.  The serial path adds inflation to the transfer; the
dual-core path inflates the next frame on core 1 during the transfer, so frame
time equals the transfer; the integrated path executes every stage serially.
Hatching marks the first stripe's scaling, the only scaling on the critical
path.  The dark segment is OS and interpreter residual (0.83~ms).}
\label{fig:timeline}
\end{figure*}
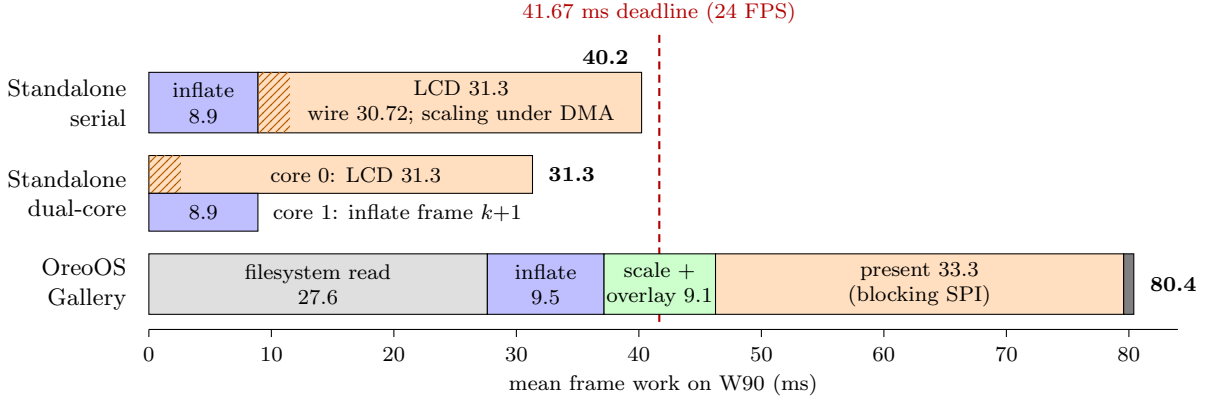

Gallery owns playback policy; at the measured revision, firmware owns only
inflation.  For each frame, Gallery reads one length-prefixed payload from the
filesystem into a reused compressed buffer and passes memoryviews of that
buffer and a reused RGB565 frame buffer to the firmware-resident
\texttt{\_oreo\_rv565} module, a MicroPython user C
module~\cite{micropython}.  That module calls
\texttt{tinfl}~\cite{miniz}, validates the exact decoded length, and returns
without constructing a temporary Python \texttt{bytes} object or a
Python-managed Deflate stream.  A separate native Gallery routine then scales
the frame into OreoOS's 153,600-byte framebuffer; Python draws a playback
overlay; and the display driver presents the frame with blocking SPI writes in
four 38,400-byte chunks separated by 200~$\mu$s pauses, which smooth supply
current on the badge.  Because the MicroPython heap resides in PSRAM on this
board, the compressed buffer, decoded frame, and framebuffer are all in
external RAM.  The stages are strictly serial (Figure~\ref{fig:timeline}).

\subsection{Scheduling and interaction semantics}

For a clip rate $f$, the period is $P=1/f$ and the nominal deadline of frame
$i$ is

\begin{equation}
d_i=t_0+iP.
\label{eq:deadline}
\end{equation}

The standalone player advances the absolute deadline of
Eq.~\eqref{eq:deadline} after every transfer and waits for the remainder of
the period.  If the clock is more than one period late, it advances both the
deadline and frame index until they again describe a future presentation;
skipped frames are counted as drops and are never decoded merely to catch up.

Gallery applies a bounded-work rule at the OS boundary instead: no more than
one frame is decoded per update tick, and accumulated delay beyond one period
is discarded rather than converted into a burst of work.  Gallery therefore
does not skip source frames when it falls behind; playback slows instead.
This bounds the time before the next input poll and makes slower playback
preferable to an unresponsive interface.

Each active-low button registers rising- and falling-edge interrupts whose
handlers only set preallocated flags; an 8~ms software debounce is applied
when the OS loop polls them.  MicroPython runs pin handlers as scheduled
(soft) callbacks unless a hard interrupt is requested~\cite{micropython2026pin},
and OreoOS uses the default, so they execute after the current native call
returns and sample the pin level at that time rather than at the edge.
Whether a tap shorter than a native frame operation is retained therefore
depends on its timing, and we make no input-latency or retention claim.

\section{Evaluation}
\label{sec:evaluation}

\begin{table}[t]
\centering
\caption{Evaluated configuration.}
\label{tab:platform}
\begin{tabular}{@{}>{\raggedright\arraybackslash}p{0.27\columnwidth}%
                >{\raggedright\arraybackslash}p{0.66\columnwidth}@{}}
\toprule
Parameter & Value \\
\midrule
Board & ESP32-S3-DevKitC-1 N16R8 \\
Processor & Dual-core Xtensa LX7, 240~MHz (configured) \\
Memory & 16~MiB flash, 8~MiB octal PSRAM \\
Display & ST7789, 320$\times$240, RGB565, 4-wire SPI; backlight 60\% PWM \\
SPI clock & 40~MHz (driver-reported) \\
Standalone & ESP-IDF v5.5.1; PSRAM allocator disabled \\
Integrated & MicroPython v1.28.0 with the firmware-resident inflater; Gallery app \\
Source & \emph{Big Buck Bunny}~\cite{blender2008bbb}, 480p H.264, CC~BY~3.0 \\
Workloads & W90: 90--100~s; W360: 360--370~s; each 240 frames, 180$\times$135, 24~FPS \\
\bottomrule
\end{tabular}
\end{table}

\subsection{Workloads and methodology}

The workloads are two ten-second windows of the openly licensed animated film
\emph{Big Buck Bunny}~\cite{blender2008bbb} (480p H.264 release, SHA-256
\texttt{b2acb9bd\ldots 55b9a27}), encoded as described above.  The windows were
chosen before measurement to differ in compressed size: W90 compresses to
6,421,314~bytes (payloads 24,341--33,261~bytes, mean 26,751) and W360 to
8,959,471~bytes (36,958--37,816~bytes, mean 37,327), against 11,664,012~bytes
for 240 raw frames.  Table~\ref{tab:platform} lists the evaluated configuration
and the commits whose sources match the recorded digests.

Timing uses the ESP32-S3 monotonic microsecond timer.  In the standalone
firmware, frame work begins when the playback loop starts a frame and ends
when that frame's final LCD transaction completes and, in the overlapped
configuration, when the next frame's inflation has been collected.
Inflation, cumulative stripe scaling, cumulative DMA wait, and the complete
LCD phase are timed separately; in the overlapped configuration, the
reported inflation time is that of the presented frame, measured on core 1.
In Gallery, frame work begins at the update tick that decodes the frame and
ends after the display flush; filesystem read, inflation, scaling, total
drawing, and presentation are timed separately, and the residual is
attributed to the OS loop and interpreter.  Displayed FPS is completed
presentations divided by wall time.  A deadline miss is frame work exceeding
the period.

Each standalone configuration uses a two-second, 48-frame warm-up and a
30-second, 720-frame timed window.  Ten independently reset runs per
configuration provide 7,200 frame samples, covering every source frame 30
times.  Frame percentiles use the pooled empirical distribution with linear
interpolation between order statistics (Hyndman and Fan's
type~7~\cite{hyndman1996sample}); run-level aggregates use two-sided 95\%
Student-$t$ intervals~\cite{student1908probable}.  Linear models are ordinary
least-squares fits, reported with Pearson's $r$.  Per-frame records are
retained in RAM and printed only after the timed window, so serial output
does not enter the reported latency.  Internal and external heap state is
sampled at boot, after the clip is opened, before playback, and after the
timed window.  The integrated experiment uses a 48-frame warm-up and one
720-frame window of W90 in the unmodified OreoOS Gallery loop.

Every result is tied to the SHA-256 digests of the application image,
media, build configuration, and relevant source files, and to the raw serial
log and per-frame CSV it was computed from.  A collector rejects any capture
whose numbered frame sequence is incomplete, and the aggregator refuses to
pool runs whose artifact hashes, configuration, or metadata differ.  The
collection and analysis scripts that derive every value in
Sections~\ref{sec:bus}--\ref{sec:gap} are included in the released
artifact~\cite{bhattacharya2026oreo}.

\subsection{Results}
\label{sec:standalone-results}

\begin{table*}[!t]
\centering
\caption{Measured \system{} v6 frame work.  Standalone rows pool ten runs of
720 frames; the integrated row is one 720-frame run.  Integrated playback does
not skip frames when late, so drops are not applicable (n/a).}
\label{tab:results}
\resizebox{\textwidth}{!}{%
\begin{tabular}{llccccccccc}
\toprule
Workload & Path & Frames & FPS & Drops & Misses & Mean (ms) & p50 (ms) & p95 (ms) & p99 (ms) & Max (ms) \\
\midrule
W90 & Standalone, serial & 7,200 & 24.002 & 0 & 0 & 40.224 & 40.205 & 40.741 & 41.348 & 41.353 \\
W90 & Standalone, dual-core overlap & 7,200 & 24.009 & 0 & 0 & 31.324 & 31.322 & 31.331 & 31.333 & 31.334 \\
W360 & Standalone, serial & 7,200 & 23.905 & 20 & 7,200 & 41.830 & 41.828 & 41.869 & 41.881 & 41.901 \\
W360 & Standalone, dual-core overlap & 7,200 & 24.009 & 0 & 0 & 31.326 & 31.324 & 31.334 & 31.335 & 31.336 \\
\midrule
W90 & OreoOS Gallery, firmware inflation & 720 & 12.170 & n/a & 720 & 80.400 & 80.174 & 86.067 & 87.117 & 92.102 \\
\bottomrule
\end{tabular}
}
\end{table*}

Table~\ref{tab:results} reports all five configurations.  The serial path
meets every 41.667~ms deadline on W90 with 0.314~ms of slack at its worst
frame, but misses every deadline on W360: its mean frame work exceeds the
period by 0.163~ms, so lateness accumulates until the deadline loop skips one
source frame roughly every 256 frames, giving 20 drops and 23.905~FPS.  The
dual-core path meets every deadline on both windows with 10.331~ms of slack
at its worst frame.  Run-level variation is negligible in every
configuration; for example, the 95\% interval on W90 overlapped mean frame
work is 31.32397--31.32401~ms.  Paced FPS is fixed by the deadline loop, so tail
margin and the miss count are the informative quantities.

The integrated path misses all 720 deadlines and presents 12.170~FPS.
Because Gallery slows rather than skips, W90 plays at 0.507$\times$ real
time there.

\subsection{Overlap removes decode from the critical path}
\label{sec:bus}

In the serial path, Eq.~\eqref{eq:work} accounts for the measured mean work
to within 2~$\mu$s on both windows (8.910 + 31.312 against 40.224~ms on W90;
10.516 + 31.312 against 41.830~ms on W360).  Mean $T_{lcd}$ is 31.312~ms
against the 30.72~ms wire time of Eq.~\eqref{eq:wire}, so the display transfer
runs at 98.1\% of the bus bound.  Within it, core 0 spends 2.598~ms scaling
and 28.470~ms waiting for DMA: it is idle for 90.9\% of the transfer, while
the content-dependent inflation that precedes it decides whether the frame is
late.

The dual-core path converts that idle time into decode capacity.  Core 1
inflates in 8.902~ms on W90 and 10.503~ms on W360, within 13~$\mu$s of the
serial path's 8.910 and 10.516~ms, so running the two cores concurrently
imposes no measurable contention penalty on inflation.  Frame work
falls to 31.324 and 31.326~ms, exceeding the transfer by 8 and 9~$\mu$s of
synchronisation, as Eq.~\eqref{eq:overlap} predicts; this is a 22.1\% and
25.1\% reduction respectively.  The schedule is now bounded by the bus alone:
inflation occupies 28--34\% of the transfer it hides behind, and the
remaining headroom is 10.3~ms per frame at 24~FPS.  Paced at 24~FPS, the
deadline loop does not exercise that headroom; the transfer time implies a
ceiling of about 31.9~FPS for full-screen frames, which we did not measure.

\subsection{Decode cost is predictable from the container}
\label{sec:model}

Inflation time is linear in the stored payload length $L$.  On W90, whose
payloads span 24,341--33,261~bytes, the serial runs give
$T_{inflate}\approx4.950~\text{ms}+0.1480~\mu\text{s/byte}\times L$ with
$r^2=0.987$; fitting on even-numbered source frames and testing on
odd-numbered frames gives a mean absolute error of 0.023~ms.  W360's payloads
span only 2.3\%, so its own fit is poorly conditioned ($r^2=0.200$) even
though its held-out error is 0.016~ms.  The informative test is therefore
transfer between windows: the W90 model predicts every W360 frame with a mean
error of 0.040~ms (maximum 0.097~ms), whereas a model fitted on W360's narrow
range alone mispredicts W90 by 0.86~ms on average.  Pooling both windows gives

\begin{equation}
T_{inflate}\approx 4.858\ \text{ms} + 0.1515\ \mu\text{s/byte}\times L,
\label{eq:inflate}
\end{equation}

with $r^2=0.9987$ and a mean absolute error of 0.022~ms.  The intercept
reflects the fixed cost of producing 48,600 output bytes; the slope reflects
symbol decoding.

Because $L$ is stored in the container, schedulability is decidable at
ingest.  Substituting Eq.~\eqref{eq:inflate} and the measured $T_{lcd}$ into
Eq.~\eqref{eq:work}, the serial path meets the period $P$ only for frames with

\begin{equation}
L \le L^{*}=\frac{P-T_{lcd}-a}{b}\approx 36{,}273\ \text{bytes},
\label{eq:threshold}
\end{equation}

where $a$ and $b$ are the intercept and slope of Eq.~\eqref{eq:inflate}.
W90's largest payload is 33,261~bytes and W360's smallest is 36,958~bytes, so
the threshold predicts that no W90 frame and every W360 frame is late; the
measurements agree on all 480 source frames.  Under Eq.~\eqref{eq:overlap},
inflation stays hidden while $a+bL\le T_{lcd}$, that is, for payloads up to
about 175~KB, far above the 48,600-byte raw frame; within the measured range,
the dual-core path's schedulability no longer depends on content.  This property follows from
independent, single-codec frames; it would not hold for formats whose decode
cost depends on reference state.

\subsection{The cost of frame independence}
\label{sec:independence}

To measure what independence gives up, we decoded each evaluated container
losslessly on the host and recompressed the identical RGB565 frames under
alternative organisations (Table~\ref{tab:independence}).  Recompressing each
frame at level~1 reproduces all 480 shipped payloads byte for byte,
confirming the setup.

\begin{table}[t]
\centering
\caption{Deflate payload bytes for the evaluated frames under alternative
organisations, relative to the shipped independent level-1 payloads of the
same window.}
\label{tab:independence}
\resizebox{\columnwidth}{!}{%
\begin{tabular}{lrrrr}
\toprule
 & \multicolumn{2}{c}{W90} & \multicolumn{2}{c}{W360} \\
Organisation & Bytes & Rel. & Bytes & Rel. \\
\midrule
Independent, level 1 (shipped) & 6,420,342 & 1.000 & 8,958,499 & 1.000 \\
Single stream, level 1 & 6,488,284 & 1.011 & 8,947,582 & 0.999 \\
Independent, level 6 & 6,027,250 & 0.939 & 8,881,173 & 0.991 \\
Independent, level 9 & 6,024,303 & 0.938 & 8,881,230 & 0.991 \\
Single stream, level 9 & 6,076,028 & 0.946 & 8,864,882 & 0.990 \\
XOR with previous frame, level 1 & 1,254,138 & 0.195 & 4,840,026 & 0.540 \\
\bottomrule
\end{tabular}
}
\end{table}

Within Deflate, independence is free, as Eq.~\eqref{eq:window} predicts for a
48,600-byte frame.  A single shared stream is 1.1\% \emph{larger} than
independent frames on W90 and 0.1\% smaller on W360 at level~1, and differs
by at most 0.9\% at level~9: no co-located pixel of the previous frame is
reachable.  The compression level matters more on W90 (6.1\% from level~1 to
level~6) than on W360 (0.9\%); its effect on device inflation time was not
measured, so level~1 remains the evaluated configuration.

Independence is not free relative to explicit temporal prediction.  XOR
against the previous frame, followed by the same level-1 Deflate, shrinks the
payload by 80.5\% on W90 and 46.0\% on W360: the animation contains large
static regions whose redundancy lies exactly $2WH$ bytes away, out of
Deflate's reach.  The true price of independence on this content is therefore
a 1.9--5.1$\times$ larger stream, paid in exchange for $O(1)$ seeking, local
error recovery, and the length-predictable decode cost of
Section~\ref{sec:model}.  A keyframe-plus-delta variant with a bounded group
of frames would recover part of that price; with the second decoded buffer the
dual-core path already holds a reference frame, but its decode cost would then
depend on the reference chain.

\subsection{Anatomy of the integration gap}
\label{sec:gap}

Table~\ref{tab:gallery} decomposes integrated frame work on W90.  The
interpreter and OS loop account for only 0.83~ms on average (p99 1.32~ms).
The gap to the standalone path comes from three structural differences.

\begin{table}[t]
\centering
\caption{Integrated Gallery stage times on W90 (720 frames).  Overlay is
drawing time other than scaling; the residual is OS loop and interpreter
work.}
\label{tab:gallery}
\small
\setlength{\tabcolsep}{4pt}
\begin{tabular}{lrrr}
\toprule
Stage & Mean (ms) & p99 (ms) & Max (ms) \\
\midrule
Filesystem read & 27.631 & 33.546 & 34.383 \\
Inflation & 9.525 & 10.574 & 10.850 \\
Scaling (native) & 7.641 & 8.194 & 8.743 \\
Overlay & 1.459 & 2.074 & 7.010 \\
Presentation & 33.317 & 33.918 & 34.344 \\
OS residual & 0.827 & 1.315 & 5.204 \\
\midrule
Frame work & 80.400 & 87.117 & 92.102 \\
\bottomrule
\end{tabular}
\end{table}

\paragraph{Filesystem input.}  Reading the mean 26,751-byte payload takes
27.631~ms, an effective throughput of 0.97~MB/s.  Scaled proportionally to an
uncompressed 48,600-byte frame, the read alone would take about 50.2~ms,
beyond the 41.7~ms period.  On this path, Deflate is therefore not merely a
storage optimisation: it is what brings input below the period at all.  The
standalone path avoids the cost entirely by reading memory-mapped flash.

\paragraph{Memory placement.}  Integrated inflation takes 9.525~ms against
8.910~ms standalone on the same frames.  Its per-byte cost is similar (0.1416
against 0.1480~$\mu$s/byte) but its fixed cost is 0.79~ms higher (5.738
against 4.950~ms), consistent with writing the 48,600-byte output to PSRAM
instead of internal SRAM.  Scaling takes 7.641~ms against 2.598~ms; it writes
a complete 153,600-byte framebuffer in PSRAM rather than SRAM stripes, though
the two kernels also differ, so the 2.9$\times$ ratio is not attributed to
memory placement alone.

\paragraph{Serialized presentation.}  Presentation costs 33.317~ms, 2.60~ms
above the wire time, of which 0.6~ms are the deliberate inter-chunk pauses.
Unlike the standalone paths, no other stage runs during it.

These results determine the next step.  Removing the filesystem read alone
would leave about 52.8~ms of serial work, still above the period.  Meeting
24~FPS inside OreoOS requires the standalone structure---memory-resident
input, SRAM stripes, scaling overlapped with DMA, and inflation on the second
core---to become a single firmware operation invoked once per frame by
Gallery.

\subsection{Memory}

Table~\ref{tab:resources} reports internal-SRAM allocation for both
standalone configurations, identical across all runs and both windows.  The
serial path allocates 81,816~bytes beyond the clip tables: the 48,600-byte
decoded frame, two 15,360-byte stripes, and allocator overhead.  The
dual-core path allocates 156,096~bytes; the 74,280-byte difference is the
second decoded frame, the 24~KiB worker stack, and the queues and task
control block.  Both figures exclude the 23,040-byte per-frame measurement
array.  The 1.4~KiB drift between the pre-playback and final samples is SPI
and backlight driver initialisation, which occurs after the pre-playback
sample; it is identical in every run, so playback does not grow the heap.
Integrated memory was not sampled during playback; by construction, Gallery
allocates a 48,600-byte frame buffer and a 49,624-byte compressed buffer in
addition to the 153,600-byte framebuffer that OreoOS already holds, all in
PSRAM.

\begin{table}[t]
\centering
\caption{Measured standalone memory use (internal SRAM; identical in all ten
runs of both windows).}
\label{tab:resources}
\resizebox{\columnwidth}{!}{%
\begin{tabular}{lrrrr}
\toprule
Configuration & SRAM & PSRAM & Drift & Minimum free \\
\midrule
Serial & 79.9 KiB & 0 KiB & 1.4 KiB & 240.4 KiB \\
Dual-core overlap & 152.4 KiB & 0 KiB & 1.4 KiB & 167.9 KiB \\
\bottomrule
\end{tabular}
}
\end{table}

\section{Limitations and Threats to Validity}

The evaluation uses two ten-second windows of one animated film, one board,
and one resolution and rate.  The coefficients of Eq.~\eqref{eq:inflate} and
the independence measurements are properties of this content and
configuration; footage with different statistics, such as camera noise, may
change them, although Eq.~\eqref{eq:window} does not depend on content.  The
decode model was validated across the two windows, not across films or
resolutions, and the dual-core condition beyond the measured payload range is
an extrapolation.  We report no image-quality metric: loss arises only from
resampling and RGB565 quantisation, which an evaluation must measure against
a fixed resized RGB888 reference using PSNR and SSIM~\cite{wang2004image}.

No codec baseline is reported.  A comparison with MJPEG or with Espressif's
player~\cite{espressif2026player} must hold decoded dimensions, rate, clip
range, and quality constant, and must consider the complete resource envelope
rather than size or FPS alone.  The serial-against-overlap comparison is a
controlled ablation, but the integrated row differs from the standalone rows
in input path, memory placement, scaling kernel, presentation, and the
presence of the OS; Section~\ref{sec:gap} separates these by stage timing
rather than by controlled removal.  The integrated path does not yet use the
dual-core pipeline.  Input latency was not measured.  The CPU clock is the
configured value, not a measurement.

\system{} is currently silent, RGB565-specific, and intended for small
displays.  Future work includes the fused dual-core firmware path for OreoOS,
a keyframe-plus-delta variant that keeps decode cost bounded, a
schedulability check in the ingest tool based on Eq.~\eqref{eq:threshold},
SD-card streaming, synchronised audio, quality-matched codec baselines,
logic-analyser input-latency measurement, and other processors and bus
widths.

\section{Conclusion}

\system{} treats microcontroller video as one timed storage, memory,
decoding, scheduling, and display problem.  On an ESP32-S3, its dual-core
pipeline hides inflation behind the display transfer and sustains 24~FPS with
zero misses and 10.3~ms of slack on content that a serial path cannot
schedule, leaving the frame bounded only by an SPI transfer that runs at
98.1\% of its wire-time limit.  Independent Deflate frames make decode cost a
linear function of a length the container already stores, so schedulability
can be decided before playback; above 128$\times$128 pixels they cost nothing
relative to a shared Deflate stream, while giving up the 46--80\% that
explicit temporal prediction would save.  Inside a MicroPython operating
system, moving inflation into firmware is insufficient: the stage-resolved
measurements place the remaining cost in serialized filesystem input,
external-memory placement, and blocking presentation rather than in the
interpreter, and identify the fused firmware operation that should come next.

\section*{Artifact Availability}

OreoOS, the \system{} encoders, firmware modules, standalone benchmark, paper
source, raw serial logs, per-frame data, artifact hashes, the encoded
evaluation media, and the scripts that produce every reported value are
archived at \url{https://doi.org/\artifactdoi}~\cite{bhattacharya2026oreo}.

\begingroup
\fontsize{7pt}{7.4pt}\selectfont
\bibliographystyle{unsrt}
\bibliography{references}
\endgroup

\end{document}